\documentclass[a4paper,11pt,twoside]{article}
\usepackage{amsmath,amssymb,amsfonts,amsthm,graphicx}
\usepackage[sans]{dsfont}
\usepackage[mathscr]{euscript}
\usepackage{cite}
\usepackage[british]{babel}

\usepackage[all]{xy}

\newcommand{\rd}{\mathrm{d}}
\newcommand{\ri}{\mathrm{i}}
\newcommand{\re}{\mathrm{e}}
\newcommand{\openone}{\mathds1}
\newcommand{\norm}[1]{\left\Vert#1\right\Vert}
\newcommand{\abs}[1]{\left\vert#1\right\vert}

\newcommand{\Rbb}{\mathbb{R}}
\newcommand{\Cbb}{\mathbb{C}}
\newcommand{\Pbb}{\mathbb{P}}
\newcommand{\Qbb}{\mathbb{Q}}
\newcommand{\Ebb}{\operatorname{\mathbb{E}}}

\newcommand{\Var}{\operatorname{Var}}
\newcommand{\Real}{\operatorname{Re}}
\newcommand{\Tr}{\operatorname{Tr}}

\newcommand{\Acal}{\mathcal{A}}

 \newcommand{\Dcal}{\mathcal{D}}

\newcommand{\Ical}{\mathcal{I}}

\newcommand{\Lcal}{\mathcal{L}}

 \newcommand{\Escr}{\mathscr{E}}
\newcommand{\Fscr}{\mathscr{F}}

\newcommand{\Hscr}{\mathscr{H}}

\newcommand{\Lscr}{\mathscr{L}}

\newcommand{\Sscr}{\mathscr{S}}
\newcommand{\Tscr}{\mathscr{T}}

\begin{document}
\title{Quantum measurements in continuous time, \\ non
Markovian evolutions and feedback}

\author{Alberto Barchielli and Matteo Gregoratti
\\
Politecnico di Milano, Dipartimento di Matematica,\\
Piazza Leonardo da Vinci 33, I-20133 Milano, Italy.\\
Also: Istituto Nazionale di Fisica Nucleare, Sezione di Milano.}
\markboth{Quantum trajectories and feedback}{A. Barchielli and M. Gregoratti}
\maketitle
\abstract{
In this article we reconsider a version of quantum trajectory theory based on the stochastic Schr\"odinger equation with stochastic coefficients, which was mathematically introduced in the '90s, and we develop it in order to describe the non Markovian evolution of a quantum system continuously measured and controlled thanks to a measurement based feedback. Indeed, realistic descriptions of a feedback loop have to include delay and thus need a non Markovian theory. The theory allows to put together non Markovian evolutions and measurements in continuous time in agreement with the modern axiomatic formulation of quantum mechanics.
To illustrate the possibilities of such a theory, we apply it to a two-level atom stimulated by a laser. We introduce closed loop control too, via the stimulating laser, with the aim to enhance the ``squeezing'' of the emitted light, or other typical quantum properties. Note that here we change the point of view with respect to the usual applications of control theory. In our model the ``system'' is the two-level atom, but we do not want to control its state, to bring the atom to a final target state. Our aim is to control the ``Mandel $Q$-parameter'' and the spectrum of the emitted light; in particular the spectrum is not a property at a single time, but involves a long interval of times (a Fourier transform of the autocorrelation function of the observed output is needed).
}

\section{Quantum trajectories and control}

Stochastic wave function methods for the description of open quantum systems are now widely used
\cite{Car93, ZolG97, BreP02, GarZ04, Car08} and are often referred to as \emph{quantum trajectory theory}. These approaches are very important for numerical simulations and allow the continuous measurement description of detection schemes in quantum optics, namely direct, homodyne and heterodyne photo-detection \cite{BarB91, Bar90, Hol01, BarG09}. In the Markovian case, the stochastic differential equations of the quantum trajectory theory can be interpreted in terms of measurements in continuous time because they can be related to \emph{positive operator valued measures} and \emph{instruments} \cite{DavL70,Dav76,Oza84}, which are the objects representing observables and state changes in the modern axiomatic formulation of quantum mechanics. Moreover, these stochastic differential equations can be deduced from purely quantum evolution equations for the measured system coupled with a quantum environment, combined with a continuous monitoring of the environment itself. Such a representation is based on the use of quantum fields and quantum stochastic calculus \cite{BarL85, Bar86, Bel88, Bel89, BarP96, Hol01, Bar06}.

The whole quantum trajectory theory is well developed in the Markovian case, but to include memory effects is more and more important. A generalization by Di\'osi, Gisin and Strunz, based on the introduction of functional derivatives acting on the ``past'' inside the \emph{stochastic Schr\"odinger equation} (SSE), is able to describe dynamical memory effects \cite{DioS97, DioGS98}, but fails to have an interpretation in terms of continuous measurements \cite{WisG08}.

Another way to include non Markovian effects and to maintain at the same time the continuous measurement interpretation is to start from the linear SSE and to generalize it by allowing for stochastic coefficients. This can be done without violating the axiomatic formulation of quantum mechanics and a non Markovian quantum trajectory theory can be developed in a mathematically consistent way \cite{BarH95}. The key point to get this result is that the non Markovian character is introduced by using stochastic coefficients in the linear version of the stochastic Schr\"odinger equation, without other modifications, and this allows to construct completely positive linear dynamics and instruments as required by the principles of a quantum theory.

More recently, motivated by the growing interest for non Markovian evolutions, we begun to analyse possible physical applications of this theory. Some applications to systems affected by coloured noises and continuously monitored have been already developed \cite{BarPP10, BarDPP11}. This paper, instead, will be focused mainly on feedback control.

In quantum optical systems, even when the Markov approximation for the reduced dynamics is well justified, memory can enter into play when imperfections in the stimulating lasers are taken into account
\cite{BarP02} and when feedback loops are introduced to control the system \cite{Bel83, WisM93, WisM94, GTV99, WanW01, WanWM01, GarZ04, NKI09,WisM10}. The so called \emph{closed loop control} is based on the continuous monitoring of the system and, so, it fits well in the theory of measurements in continuous time. In some approximations, one can consider an instantaneous and very singular feedback and in this case the usual Markov framework is sufficient; however, more realistic descriptions of the feedback loop, including delay, need a non Markovian theory \cite{WisM93, WisM94, GTV99,BarG09,WanW01,WanWM01,NKI09,WisM10}.

In this paper we present the non Markovian version of the theory of quantum measurements in continuous time, based on the SSE and the \emph{stochastic master equation} (SME) with stochastic coefficients. Then, we develop such a theory, we explain how the stochastic coefficients can describe the most general measurement based feedback and also some imperfections in the apparatus, and we show how to get the physical probabilities for the output of the observation, its moments and other related quantities of physical interest. Indeed, in quantum optical systems the moments of the stochastic output are connected to the Mandel $Q$-parameter and to the spectrum of the emitted light (homodyne and heterodyne spectra) and allow for the study of typical quantum properties of the emitted light, such as \emph{squeezing} \cite{BarG08, BarG09, BarGL09}. To illustrate these concepts we shall use a prototype model, a two-level atom stimulated by a laser, which is known to have a rich spectrum and to emit squeezed light under particular conditions. Our emphasis will be on the possibility of modelling a non perfectly monochromatic and coherent stimulating laser and of modelling a measurement based feedback. We shall introduce closed loop control, via the stimulating laser, with the aim to control the squeezing in the observed spectrum, not to control the state of the system. Let us stress the change of point of view with respect to the usual applications of control theory \cite{BelE08,BouVH08,Gough08,James08}. Here the ``system'' is the two-level atom, but we do not want to control its state, so as to bring the atom to a final target state, which is however possible inside our theory. Our aim is to control the properties of the emitted light; moreover, we want to control the spectrum, which is not a property at a single time, but involves a long interval of times (a Fourier transform in time is needed).

\section{The stochastic Schr\"odinger equation and the stochastic master equation}\label{sec:SME}

The best way to introduce memory in quantum evolutions is to start from a dynamical equation in Hilbert space; this approach automatically guarantees the complete positivity of the evolution of the state (statistical operator) of the system. Moreover, considering the linear version of the SSE allows to construct the \emph{instruments} related to the continuous monitoring even in the non Markov case \cite{BarH95, BarPP10, BarDPP11}. We shall introduce first several mathematical objects, from the linear SSE \eqref{lSSE} to the instruments \eqref{instruments}, and later, thanks to these latter, we shall give a consistent physical interpretation of the whole construction.

Let $\Hscr$ be the Hilbert space of the quantum system of interest, a separable complex Hilbert space, with inner product $\langle \cdot |\cdot \rangle$. Moreover, we denote by $\Lscr(\Hscr)$ the space of the bounded operators on $\Hscr$, by $\Tscr(\Hscr)\subset \Lscr(\Hscr)$ the trace class and by $\Sscr(\Hscr)\subset \Tscr(\Hscr)$ the convex set of the statistical operators.

\subsection{The linear SSE and the reference probability}
Let us consider a reference probability space $( \Omega, \Fscr, \Qbb)$ with a filtration of $\sigma$-algebras $(\Fscr_t)_t$ satisfying the usual hypotheses, i.e.\ $A\in\Fscr$ with $\Qbb(A)=0$ implies $A\in \Fscr_0$, and $\Fscr_t=\bigcap_{T>t}\Fscr_T$. As usual, we denote by $\omega$ the generic sample point in the sample space $\Omega$. The $d+d'$ driving noises of the key SDE \eqref{lSSE} are defined in this filtered probability space; they are $d$ continuous standard Wiener processes $B_1, \ldots, B_d$ and $d'$ Poisson processes $N_{d+1},\ldots,N_{d+d'}$. Under the reference probability $\Qbb$, all these processes are independent and are adapted, with increments independent from the past, with respect to the given filtration. Every Poisson process $N_k$ is taken with trajectories continuous from the right and with limits from the left (c\`{a}dl\`{a}g); let $\lambda_k>0$ be the intensity of $N_k$.

We assume also we have a set of stochastic processes $H(t)$, $L_i(t)\ (i=1,\ldots, d+d')$,  with values in $\Lscr(\Hscr)$, such that $ H(t,\omega)^* =H(t,\omega)$ and $t\mapsto L_i(t) $, $t\mapsto H(t) $ are adapted processes with trajectories continuous from the left and with limits from the right  (c\`{a}gl\`{a}d, continuity in the strong operator topology). We assume also
\[
\int_0^T\biggl(\norm{H(t)}+\sum_{i=1}^{d+d'}\norm{L_i(t)}^2 \biggr)\rd t\leq M(T)<+\infty, \qquad \forall T>0,
\]
where $M(T)$ does not depend on the sample point $\omega$.

Then, we introduce the linear SSE, for processes with values in $\Hscr$,
\begin{multline}\label{lSSE}
\rd  \phi(t)= \biggl[-\ri  H(t) -\frac{1}2\sum_{i=1}^{d+d'} L_i(t)^*  L_i(t)+\frac 1 2\sum_{k=d+1}^{d+d'} \lambda_k \biggr]\phi(t_-) \rd  t\\{} + \sum_{i=1}^d L_i(t)\phi(t_-)\, \rd  B_i(t)
+ \sum_{k=d+1}^{d+d'} \left(\frac{L_k(t)}{\sqrt{\lambda_k}}-\openone\right)\phi(t_-)\,\rd  N_k(t),
\end{multline}
with initial condition
\begin{equation}\label{phi_0}
\phi(0)=\phi_0\in \Hscr, \qquad \norm{\phi_0} =1.
\end{equation}
Equation \eqref{lSSE} is an It\^o-type stochastic differential equation
admitting a unique strong solution \cite[Proposition 2.1]{BarH95}. The solution is continuous from the right and with limits from the left (c\`{a}dl\`{a}g); the symbol $t_-$ means to take the limit from the left.

Here only bounded coefficients are considered, in order not to have mathematical complications, but generalizations to unbounded coefficients are of physical interest. Note that the filtration $(\Fscr_t)$ can be taken bigger than the natural filtration generated by the driving noises $B$ and $N$, so that the processes $L_i,\,H$ can depend also on some other external noises.

\subsection{The linear stochastic master equation}\label{sec:lSME}

The SSE \eqref{lSSE} can be translated into a stochastic differential equation for trace class operators. Indeed, adopting for a moment the Dirac notation, we can consider the rank one selfadjoint operator $|\phi(t)\rangle \langle \phi(t)|$ and compute its stochastic differential. In this way we get a closed linear equation, which can be extended to a process $\sigma(t)$ with values in $\Tscr(\Hscr)$ and whose initial condition can be any pure or mixed state. This is the linear stochastic master equation (SME) \cite[Propositions 3.2 and 3.4]{BarH95}:
\begin{multline}\label{lSME}
\rd \sigma(t)=\Lcal(t)[  \sigma(t_-)]\rd  t+ \sum_{i=1}^d \bigl( L_i(t)
  \sigma(t_-) +   \sigma(t_-) L_i(t)^* \bigr)\rd
B_i(t)\\ {}+\sum_{k=d+1}^{d'}\left(\frac 1 {\lambda_k}\,L_k(t)   \sigma(t_-) L_k(t)^*
-  \sigma(t_-)\right)\bigl( \rd  N_k(t) -\lambda_k\rd  t \bigr),
\end{multline}
with initial condition
\begin{equation}\label{rho_0}
\sigma(0)=\rho_0\in \Sscr(\Hscr).
\end{equation}
The operator $\Lcal(t)$ is the stochastic Liouvillian
\begin{equation}\label{randomLcal}
\Lcal(t)[\tau]:=-\ri  \left[H(t),\, \tau \right] -\frac 1 2 \sum_{i=1}^{d+d'}
\left\{L_i(t)^*  L_i(t),\, \tau\right\}
+ \sum_{i=1}^{d+d'}  L_i(t) \tau L_i(t)^* .
\end{equation}
The SME \eqref{lSME} admits a unique strong solution. Typically, the solution $\sigma(t)$ is not Markovian because $\Lcal(t)$ depends on the past through the random operators $H(t)$, $L_i(t)$.

\paragraph{The propagator.} In the following we shall need the fundamental solution, or \emph{propagator}, $\Acal(t,s)$ of Eq.\ \eqref{lSME}, i.e.\ the random linear map on $\Tscr(\Hscr)$ defined by $\sigma(s)\mapsto \sigma(t)$. By construction $\Acal(t,s)$ is completely positive and satisfies the composition law $\Acal(t,s)\circ \Acal(s,r) = \Acal (t,r)$ for $0\leq r\leq s \leq t$. The propagator $\Acal(t,s)$ solves Eq.\ \eqref{lSME} with initial condition $\Acal(s,s)=\openone$.

\subsection{The new probability and the non linear equations}
Let us fix a non random state $z\in \Sscr(\Hscr)$  and define the stochastic processes
\begin{equation}\label{def:aposts}
p(t):= \Tr\{\sigma(t)\}, \qquad \rho(t,\omega):=\begin{cases} p(t,\omega)^{-1} \sigma(t,\omega) & \text{if } p(t,\omega)\neq 0,
\\
z & \text{if } p(t,\omega)= 0, \end{cases}
\end{equation}
\begin{subequations}\label{miik}
\begin{equation}\label{mi}
m_i(t):=2\Real\Tr\left\{L_i(t)\rho(t_-)\right\} , \qquad i=1,\ldots,d,
\end{equation}
\begin{equation}\label{ik}
i_k(t):=\Tr\left\{L_k(t)^*L_k(t)\rho(t_-)\right\}, \qquad k=d+1,\ldots,d+d'.
\end{equation}
\end{subequations}
By taking the trace of \eqref{lSME}, we have that $p(t)$ satisfies the Dol\'{e}ans equation
\begin{equation}\label{pEq}
\rd  p(t)= p(t_-)\left\{\sum_{i=1}^d m_i(t) \rd  B_i(t)+ \sum_{k=d+1}^{d+d'}\left( \frac{i_k(t)}{\lambda_k}-
1\right)\bigl(\rd  N_k (t)-\lambda_k\rd  t\bigr)\right\}
\end{equation}
with $p(0)=1$ and where the coefficients $m_i(t)$ and $i_k(t)$ depend on the initial condition $\rho_0$ \eqref{rho_0}.

\subsubsection{The new probability}
The key property of quantum trajectory theory is that Eq.\ \eqref{pEq} implies that $p(t)$ is a mean-one $\Qbb$-martingale \cite[Theorem 2.4, Section 3.1]{BarH95}. This allows us to define the new probabilities
\begin{equation}\label{newprob}
\Pbb^T_{\rho_0}(F):=\int_F p(T,\omega)\,\Qbb(\rd\omega)=\Ebb_\Qbb[p(T)1_F], \quad \forall F\in \Fscr_T.
\end{equation}
Due to the martingale property of $p(t)$, the probabilities $\Pbb^T_{\rho_0}$ are consistent, in the sense that $\Pbb^t_{\rho_0}(F)=\Pbb^s_{\rho_0}(F)$ for $F\in \Fscr_s$, $t\geq s \geq 0$.

Of course, the new probability $\Pbb^T_{\rho_0}$ modifies the distributions of all the processes appearing in the theory, in particular the distribution of the processes $B_i$ and $N_k$.

\subsubsection{The Girsanov transformation}
A very important property is that a Girsanov-type theorem holds \cite[Proposition 2.5, Remarks 2.6 and 3.5]{BarH95}. Under $\Pbb^T_{\rho_0}$, in the time interval $[0,T]$, the processes
\begin{equation}
W_j(t):= B_j(t)-\int_0^t m_j(s)\rd  s, \qquad j=1,\ldots, d,
\end{equation}
turn out to be independent Wiener processes. Moreover, $N_{d+1},\ldots,N_{d+d'}$ become  simple regular c\`{a}dl\`{a}g counting process of stochastic intensities $i_{d+1},\ldots,i_{d+d'}$; the meaning of stochastic intensity is given by the heuristic formula
\begin{equation*}
\Ebb_{\Pbb^T_{\rho_0}}[\rd N_k(t)|\Fscr_t]=i_k(t)\rd t.
\end{equation*}

From these results we have immediately
\[
\frac{\rd \ }{\rd t}\,\Ebb_{\Pbb^T_{\rho_0}}\left[B_i(t)\right] =\Ebb_{\Pbb^T_{\rho_0}}\left[m_i(t)\right]
=2\Real \Ebb_{\Qbb}\left[\Tr\left\{L_i(t)\sigma(t_-)\right\}\right],
\]
\[
\frac{\rd \ }{\rd t}\,\Ebb_{\Pbb^T_{\rho_0}}\left[N_k(t)\right] =\Ebb_{\Pbb^T_{\rho_0}}\left[i_k(t)\right]
=\Ebb_{\Qbb}\left[\Tr\left\{L_k(t)^*L_k(t)\sigma(t_-)\right\}\right].
\]

\subsubsection{The non linear SSE}
In this subsection let us consider again a pure initial condition $\rho_0=|\phi_0\rangle \langle \phi_0|$ \eqref{phi_0}. In this case we have $\sigma(t)=|\phi(t)\rangle \langle \phi(t)|$ and, so,
\[
p(t)=\norm{\phi(t)}^2, \qquad m_i(t)=2\Real\langle \psi(t_-)|L_i(t)\psi(t_-)\rangle , \quad i=1,\ldots,d,
\]
\[
i_k(t)=\norm{L_k(t)\psi(t_-)}^2, \qquad k=d+1,\ldots,d+d',
\]
where $\psi(t) $ is defined by normalizing the random vector $\phi(t)$, solution of the linear SSE \eqref{lSSE}:
\[
\psi(t,\omega):=\begin{cases} \norm{\phi(t,\omega)}^{-1} \phi(t,\omega) & \text{if } \norm{\phi(t,\omega)}\neq 0,
\\
\phi_0 & \text{if } \norm{\phi(t,\omega)}= 0. \end{cases}
\]

Under the new probability $\Pbb^T_{\rho_0}$ \eqref{newprob}, the normalized Hilbert space process $\psi(t)$ turns out to satisfy the non linear stochastic equation \cite[Theorem 2.7]{BarH95}
\begin{multline}\label{nlSSE}
\rd \psi(t) = \hat K(t)\psi(t_-) \rd t + \sum_{i=1}^d \left(
L_{i}(t) -\frac 1 2 \, m_i(t)\right)\psi(t_-) \rd W_{i}(t)
\\ {}+\sum_{k=d+1}^{d+d'} \left(\frac {L_k(t)\psi(t_-)} {\sqrt{i_k(t)}}
-\psi(t_-)\right)\rd N_k(t),
\end{multline}
with initial condition $\psi(0)=\phi_0$ and
\[
\hat K(t):=-\ri H(t) - \frac 1 2 \sum_{i=1}^{d+d'} L_i(t)^* L_i(t)
-\frac 1 8 \sum_{i=1}^d m_i(t)^2+ \frac 1 2 \sum_{i=1}^d m_i(t)L_i(t) + \frac
12\sum_{k=d+1}^{d+d'} i_k(t) .
\]
It is equation \eqref{nlSSE} which is usually called SSE and which is the starting point for powerful numerical simulations.

\subsubsection{The non linear SME}
Going back to the case of a generic initial state \eqref{rho_0}, it is possible to show that the stochastic state $\rho(t)$ defined by \eqref{def:aposts} satisfies a non linear SME under the new probability $\Pbb^T_{\rho_0}$ \cite[Remark 3.6]{BarH95}:
\begin{multline}\label{nlSME}
\rd \rho(t)=\Lcal(t)[  \rho(t_-)]\rd  t+ \sum_{i=1}^d \left( L_i(t)
\rho(t_-) +   \rho(t_-) L_i(t)^* -m_i(t)\rho(t)\right)\rd
W_i(t)\\ {}+\sum_{k=d+1}^{d'}\left(\frac 1 {i_k(t)}\,L_k(t)   \rho(t_-) L_k(t)^*
-  \rho(t_-)\right)\bigl( \rd  N_k(t) -i_k(t)\rd  t \bigr),
\end{multline}
$\rho(0)=\rho_0 \in \Sscr(\Hscr)$; the operator $\Lcal(t)$ is the stochastic Liouvillian
\eqref{randomLcal}. In \cite{BarG09} the reader can find a complete discussion of the relations among the four stochastic differential equations \eqref{lSSE}, \eqref{lSME}, \eqref{nlSSE}, \eqref{nlSME}, in the purely diffusive case with deterministic coefficients.

\subsection{The continuous measurement process}\label{cont_meas}
Let us introduce now the real processes
\begin{subequations}\label{outputs}
\begin{equation}\label{Jl}
J_\ell(t):= \sum_{j=1}^d\int_0^t a_{\ell j}(t-s)\,\rd  B_j(s)+ e_\ell(t), \qquad \ell=1,\ldots,m_J,
\end{equation}
\begin{equation}\label{Ih}
I_h(t):= \sum_{k=d+1}^{d+d'}\int_{(0,t]} a_{hk}(t-s)\,\rd  N_k(s)+w_h(t), \qquad h=m_J+1,\ldots,m_I+m_J,
\end{equation}
\end{subequations}
where the integral kernels $a_{\ell j}(t-s)$ and $a_{hk}(t-s)$ are deterministic, and where $e_\ell(t)$ and $w_h(t)$ can be stochastic processes. Even more general expressions than the ones in Eqs.\ \eqref{outputs} could be considered. These processes will be interpreted as outputs of the continuous measurement, as explained in the next subsection. For this interpretation it is important the natural filtration generated by the processes \eqref{outputs}, which we denote by $(\Escr_t)$. We assume $\Escr_t\subseteq\Fscr_t$, $\forall t \geq 0$.

\subsubsection{Instruments and a posteriori states}\label{Instr}
Now, for $t\geq 0$, let us define the map-valued measure $\Ical_t$ \cite[Remark 4.2]{BarH95}:
\begin{equation}\label{instruments}
\Ical_t(E)[\varrho] = \int_E \Acal(t,0,\omega)[\varrho]\,\Qbb(\rd\omega), \qquad
\forall E\in \Escr_t,\quad \forall \varrho\in \Tscr(\Hscr).
\end{equation}
Such a measure has the properties:
\begin{enumerate}
\item[(i)] $\forall E\in \Escr_t$, $\Ical_t(E)$ is a completely positive linear map on $\Tscr(\Hscr)$,

\item[(ii)] $\forall \varrho\in \Tscr(\Hscr)$, $\forall a \in \Lscr(\Hscr)$, $\Tr\left\{a \Ical_t(\cdot)[\varrho]\right\}$ is $\sigma$-additive,

\item[(iii)] $\forall \varrho\in \Tscr(\Hscr)$, $\Tr\left\{\Ical_t(\Omega)[\varrho]\right\}=\Tr\left\{\varrho\right\}$.
\end{enumerate}
Such a map-valued measure $\Ical_t$ is called an \emph{instrument} with value space $(\Omega, \Escr_t)$ \cite{DavL70,Dav76,Oza84} and it can be consistently interpreted as a quantum mechanical measurement on the system $\Hscr$ of the processes $J_\ell(s)$ and $I_h(s)$ in the time interval $[0,t]$; the instrument gives both the probability distribution of the output and the state changes conditional on the observation.

According to the physical interpretation of the notion of instrument, the probability of the event $E\in \Escr_t$, when the pre-measurement state is $\rho_0$, is given by
\begin{equation}
\Tr\left\{\Ical_t(E)[\rho_0]\right\}= \Ebb_\Qbb\left[1_E\Tr\left\{\sigma(t)\right\} \right]= \Pbb_{\rho_0}^t(E)
\end{equation}
and this shows that the physical probability for the observation of the output up to time $t$ is indeed the one introduced in Eq.\ \eqref{newprob} restricted to $\Escr_t$.
When $t$ goes from 0 to $T$, the family of instruments $\Ical_t$ gives a consistent description of a continuous measurement performed on the system.

Then, Eq.\ \eqref{outputs} can be interpreted as the effect of the measuring apparatus which processes the ideal outputs $B_i(t)$ and $N_k(t)$ by the classical response functions $a_{\ell j}$ and $a_{hk}$, and which degrades the outputs by adding some more noises $e_\ell(t)$ and $w_h(t)$ due to the physical realization of the apparatus itself. In quantum optics, the typical output current of an homodyne or heterodyne detector is of the form \eqref{Jl} with
\begin{equation}\label{a_ellF}
a_{\ell j}(t)=\delta_{\ell j}F_\ell(t),
\end{equation}
where $F_\ell$ is the detector response function.  In the final part of this section we shall see how to compute some relevant properties of the outputs under the physical probability.

After all, we can say that $\big(\Escr_t\big)_{t\geq 0}$ is the filtration generated by the outputs, i.e.\ $\Escr_t$ contains only the events related to the observation of the outputs up to time $t$, while $\Fscr_t\supseteq \Escr_t$, $t\geq 0$, is the filtration generated by all the processes involved in the continuous measurement and in the evolution, from the outputs to the unobservable noises.

Now, let us take the conditional expectation on $\Escr_t$ of the random state $\rho(t)$ defined by \eqref{def:aposts}:
\begin{equation}\label{posteriorstate}
\hat \rho(t):=\Ebb_{\Pbb^t_{\rho_0}}[
\rho(t)|\Escr_t]\equiv \frac{\Ebb_{\Qbb}[  \sigma(t)|\Escr_t]}{\Tr\left\{
\Ebb_{\Qbb}[  \sigma(t)|\Escr_t]\right\}}.
\end{equation}
The interpretation is that $\hat \rho(t)$ is the conditional state one attributes to the system at time $t$ having observed the trajectory of the output up to time $t$. Indeed $\hat \rho(t)$ is $\Escr_t$-measurable, thus depending only on the trajectories of the output in $[0,t]$, and, moreover, one can directly check that \cite[Remark 4.4]{BarH95}
\begin{equation}\label{instr+post}
\int_E \hat \rho(t,\omega)\Pbb_{\rho_0}^t(\rd \omega)=\Ical_t(E)[\rho_0], \qquad \forall E\in \Escr_t.
\end{equation}
The state $\hat \rho(t)$ is the \emph{a posteriori state} at time $t$.

In the extreme case $\Escr_t=\Fscr_t$, which occurs for example when $\Fscr_t$ is generated by the processes $B_i$ and $N_k$ and just these processes are the observed output, we get $\hat \rho(t)=\rho(t)$. Therefore the evolution of the a posteriori state is completely defined by the non linear SME \eqref{nlSME} satisfied by $\rho(t)$, or, equivalently, by the SME \eqref{lSME} for $\sigma(t)$, or even, as pure states are mapped to pure states, by the SSE \eqref{lSSE} for $\phi(t)$ or the non linear SSE \eqref{nlSSE} for $\psi(t)$. Let us stress that the evolution is not Markovian due to the randomness of the operator coefficients $L_i(t)$, $H(t)$.

When $\Escr_t\subsetneqq\Fscr_t$, the a posteriori state $\hat \rho(t)$ has a non Markovian evolution which typically does not even satisfy a differential equation. In this case the SSE \eqref{lSSE} and the SME \eqref{lSME} have to be interpreted as an ideal unravelling of the physical evolution of $\hat \rho(t)$ which allows us to consistently define it, by \eqref{lSME} and \eqref{posteriorstate}, and allows us to compute, at least numerically, all the quantities of physical interest (that is to define the instruments $\Ical_t$).

The randomness of the coefficients $L_i(t)$ and $H(t)$ and the distinction between $\Escr_t$ and $\Fscr_t$ provide the SME \eqref{lSME} with a great flexibility which allows us to model non Markovian features of the dynamics due both to some environmental noises and to measurement based feedback loops.

\subsubsection{A priori states}
When the output of the continuous measurement is not taken into account, the state of the system at time $t$ is given by the mean state
\[
\eta(t):= \Ebb_{\Pbb^t_{\rho_0}}[ \hat \rho(t)]=\Ebb_\Qbb[\sigma(t)]=\Ebb_\Qbb[\Acal(t,0)[\rho_0]]=\Ical_t(\Omega)[\rho_0].
\]
The state $\eta(t)$ is the \emph{a priori state} at time $t$. Note that $\Ebb_\Qbb[\Acal(t,0)[\cdot]]$ is a completely positive, trace preserving, linear map, i.e.\ a \emph{quantum channel} in the terminology of quantum information.

From the SME \eqref{lSME} we get
\begin{equation}
\frac{\rd \ }{\rd t}\, \eta(t)=\Ebb_\Qbb\big[\Lcal(t)[\sigma(t)]\big] \equiv \int_\Omega \Lcal(t,\omega)[\sigma(t,\omega)]\Qbb(\rd \omega),
\end{equation}
which is not a closed differential equation when $\Lcal(t)$ is stochastic, contrary to the Markov case \cite[Section 3.4]{BarG09}. By the projection operator technique a closed integro-differential equation for the a priori state $\eta(t)$ could be obtained \cite{BarDPP11} (an evolution equation with memory), but this equation is too involved to be of practical use. Again \eqref{lSSE} and \eqref{lSME} are an unravelling of a non Markovian evolution. By the non linear SME \eqref{nlSME} we have also
\begin{equation*}
\frac{\rd \ }{\rd t}\, \eta(t)=\Ebb_{\Pbb^T_{\rho_0}}\big[\Lcal(t)[\rho(t)]\big] \equiv \int_\Omega \Lcal(t,\omega)[\rho(t,\omega)]\Pbb^T_{\rho_0}(\rd \omega), \qquad 0\leq t \leq T.
\end{equation*}

\subsubsection{Spectra and moments of the diffusive outputs}\label{spectra}
Let us consider an output current $J_\ell$ as given by Eq.\ \eqref{Jl}; $J_\ell$
is a diffusive stochastic process and its \emph{spectrum} is given by the classical notion \cite{How02}. If $J_\ell$ is at least asymptotically stationary and the limit in \eqref{defSP} exists in the sense of distributions in $\mu$, then the spectrum of $J_\ell$ is defined by
\begin{equation}\label{defSP}
S_\ell(\mu)=\lim_{T\to +\infty}\frac 1 T\Ebb_{\Pbb^T_{\rho_0}}\left[\abs{\int_0^T
\re^{\ri \mu t}J_\ell(t)\rd t}^2\right].
\end{equation}
In the case \eqref{a_ellF} without extra noise, $e_\ell(t)=0$, and for a detector response function going  to a Dirac delta, that is in the limit case $J_\ell=\dot B_\ell$, the spectrum becomes
\begin{equation}\label{def:spectrum}
S_\ell(\mu)=\lim_{T\to +\infty}\frac 1 T\Ebb_{\Pbb^T_{\rho_0}}\left[\abs{\int_0^T
\re^{\ri \mu t}\rd B_\ell(t)}^2\right].
\end{equation}

The spectrum depends on the distribution of the current $J_\ell(t)$, which is the output of the continuous measurement on the system, and, so,  $S_\ell$ gives information on the monitored system. Actually, the spectrum gives information also on the carrier which mediates the measurement. For example, in quantum optics, the system is typically a photoemissive source which is continuously monitored by homodyne or heterodyne detection of its emitted light. In this case, $S_\ell$ gives information both on the photoemissive source and on the fluorescence light. For example, $S_\ell<1$ reveals squeezing of the emitted light.

An expression for the autocorrelation function needed in the computation of the spectrum can be obtained by generalizing the techniques used in the Markovian case \cite[Section 4.5]{BarG09}. When the Liouville operator \eqref{randomLcal} and $B_\ell$ are independent (which implies that $B_\ell$ is not used for the feedback), we get
\begin{subequations}
\begin{equation}
\frac{\partial^2 \ \ }{\partial t\partial s}\,\Ebb_{\Pbb^T_{\rho_0}}\left[B_\ell(t)B_\ell(s)\right]
=\delta(t-s)+b_\ell(t,s) +b_\ell(s,t),
\end{equation}
\begin{multline}
b_\ell(t,s)
=1_{(0,+\infty)}(t-s)\,\Ebb_{\Qbb} \biggl[\Tr\biggl\{\left(L_\ell(t)+L_\ell(t)^*\right)\\ {}\times\Acal(t_-,s)\left[L_\ell(s)\sigma(s_-)+\sigma(s_-) L_\ell(s)^*\right]\biggr\}\biggr].
\end{multline}
\end{subequations}

Let us set $n_\ell(t)= \Ebb_{\Pbb^t_{\rho_0}}\left[m_\ell(t)\right]$ and assume that the limit $n_\infty:=
\lim_{t\to +\infty}n_\ell(t)$ exists. Then, we obtain the decomposition of the spectrum in the elastic part and the inelastic one (the spectrum of the fluctuations)
\begin{subequations}\label{gen_sp}
\begin{equation}
S_\ell(\mu)= S_{\mathrm{el}}(\mu)+ S_{\mathrm{inel}}(\mu), \qquad S_{\mathrm{el}}(\mu)= 2\pi \,n_\infty^{\;2}\, \delta(\mu),
\end{equation}
\begin{equation}
S_{\mathrm{inel}}(\mu)=1+ \lim_{T\to +\infty} \frac 2 T \int_0^T\rd t \int_0^t \rd s \, \cos \mu(t-s)\, d_\ell(t,s),
\end{equation}
\end{subequations}
\begin{multline*}
d_\ell(t,s):= b_\ell(t,s)-1_{(0,+\infty)}(t-s)\,n_\infty n_\ell(t)=
1_{(0,+\infty)}(t-s)\\ {}\times\Ebb_{\Qbb} \left[\Tr\left\{\left(L_\ell(t)+L_\ell(t)^*\right) \Acal(t_-,s)\left[L_\ell(s)\sigma(s_-)+\sigma(s_-) L_\ell(s)^*-n_\infty\sigma(s_-)\right]\right\}\right].
\end{multline*}

\subsubsection{The Mandel $Q$-parameter of the counting outputs}\label{sec:Qp}
When we consider direct detection in quantum optics, in the ideal case of noiseless counter, the output of the measurement is one of the counting processes, say $I_k=N_k$, which means to take $a_{hk}(t)=\delta_{hk}$ and $w_h(t)=0$ in Eq.\ \eqref{Ih}. In this case a typical quantity is the \emph{Mandel $Q$-parameter}, defined by
\[
Q_{k}(t;t_0) := \frac{\Var_{\Pbb^T_{\rho_0}}[N_k(t_0+t)-N_k(t_0)] }{
\Ebb_{\Pbb^T_{\rho_0}}[N_k(t_0+t)-N_k(t_0)] }-1.
\]
As for a Poisson process this parameter is zero, in quantum optics it is usual to say that in the case of a positive $Q$ parameter one has super-Poissonian light and sub-Poissonian light in the other case. Sub-Poissonian light is considered an indication of non-classical effects.

In quantum trajectory theory one can find expressions for the moments also in the counting case and we get
\[
Q_{k}(t;t_0) =\frac {V_k(t;t_0)}{M_k(t;t_0)}-M_k(t;t_0), \]
\[
M_k(t;t_0):=\Ebb_{\Pbb^T_{\rho_0}}[N_k(t_0+t)-N_k(t_0)]=\int_{t_0}^{t_0+t}\rd s \Ebb_{\Qbb}\left[\Tr\left\{L_k(s)^*L_k(s)\sigma(s_-)\right\}\right],
\]
\[
V_k(t;t_0):=\Ebb_{\Pbb^T_{\rho_0}}\left[\bigl(N_k(t_0+t)-N_k(t_0)\bigr)^2\right]-M_k(t;t_0).
\]
When $L_k(s)^*L_k(s)\Acal(s_-,r)$ is $\Qbb$-independent from $N_k(r)$, which happens when $N_k$ is not used for feedback, we get
\[
V_k(t;t_0)= 2\int_{t_0}^{t_0+t}\rd s \int_{t_0}^s \rd r
\Ebb_{\Qbb}\left[ \Tr\left\{L_k(s)^*L_k(s)\Acal(s_-,r)
\left[L_k(r)\sigma(r_-)L_k(r)^*\right]\right\}\right].
\]

\section{A paradigmatic model: the two-level atom}
As an application of the theory we consider a two-level atom stimulated by a laser; it is an ideal example, but is sufficiently rich and flexible to illustrate the possibilities of the theory. The Hilbert space is $\Cbb^2$ and the Hamiltonian part of the dynamics is given by
\begin{equation}\label{H(t)}
H(t)=H_0+H_f(t), \qquad H_0=\frac{\nu_0}2\, \sigma_z,
\qquad
H_f(t)=\overline{f(t)} \, \sigma_-
+ f(t) \sigma_+,
\end{equation}
where $\nu_0>0$ is the resonance frequency of the atom and $\sigma_z$, $\sigma_\pm$ are the usual Pauli matrices. The function $f$ is the laser wave, which can be noisy and can be controlled by the experimenter; a concrete choice for $f$ is given below.

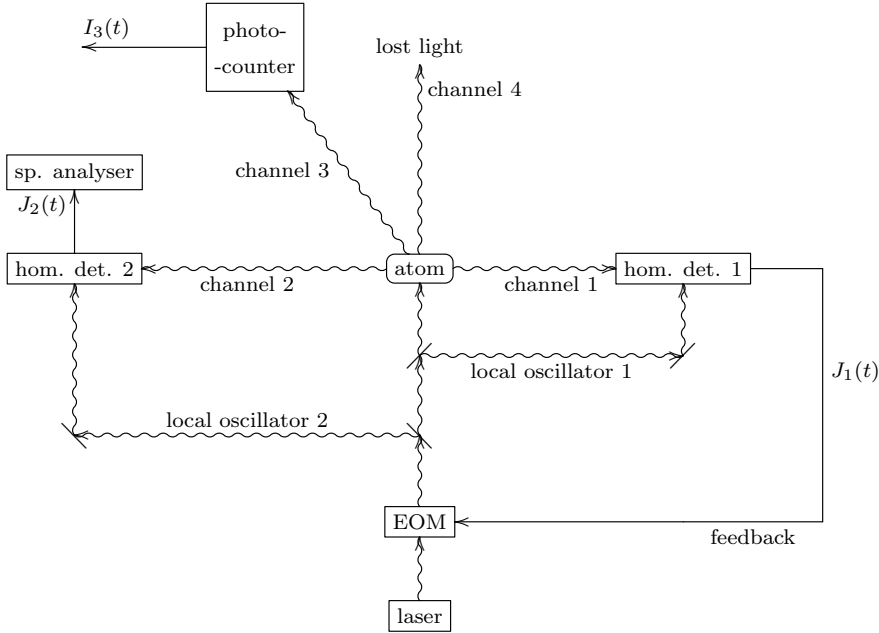
\begin{figure}[h]\begin{center}
\[
\xymatrix{ & *+[F]{\begin{matrix}\text{\begin{scriptsize}photo-\end{scriptsize}}
\\ \text{\begin{scriptsize}-counter\end{scriptsize}}\end{matrix}}\ar[l]_>>>>{ I_3(t)}& *+{\text{\begin{scriptsize}lost light\end{scriptsize}}}& & &
\\
*+[F]{\text{\begin{scriptsize}sp. analyser\end{scriptsize}}}& & &
&
&
\\
*+[F]{\text{\begin{scriptsize}hom. det. 2\end{scriptsize}}} \ar[u]^>>>{ J_2(t)}& &
*+[F-:<3pt>]{\text{\begin{scriptsize}atom\end{scriptsize}}}
\ar@{~>}[uu]_>>>>{\text{channel 4}}\ar@{~>}[rr]_{\text{channel 1}} \ar@{~>}[ll]^{\text{channel 2}} \ar@{~>}[luu]^{\text{channel 3}}& & *+[F]{\text{\begin{scriptsize}hom.
det. 1\end{scriptsize}}} \ar@{-}[r]& *=0{}\ar@{-}[ddd]^<<<<<<<<<<<<<<{ J_1(t)}
\\
&& *=0{\diagup} \ar@{~>}[u]\ar@{~>}[rr]_{\text{local oscillator
1}}& & *=0{\diagup}\ar@{~>}[u] &
\\
*=0{\diagdown}\ar@{~>}[uu]& &*=0{\diagdown} \ar@{~>}[ll]_{\text{local oscillator 2}}\ar@{~>}[u] &  & &
\\
& & *+[F]{\text{\begin{scriptsize}EOM\end{scriptsize}}}\ar@{~>}[u]&  & *=0{}\ar[ll]&*=0{}\ar@{-}[l]^{ \text{feedback}}
\\
& & *+[F]{\text{\begin{scriptsize}laser\end{scriptsize}}}\ar@{~>}[u] & & &}
\]
\caption{The electro-modulator EOM acts as a phase modulator. The beam-splitter near EOM has transmissivity and reflectivity nearly equal. The second beam-splitter has a transmissivity much smaller than the reflectivity. In this way the two local oscillators are much more intense than the light stimulating the atom, as required by the homodyne configuration.}\label{fig:feed}
\end{center}\end{figure}

Let us complete the model by choosing the noise-driven terms in the SME \eqref{lSME}, which we call channels in the sequel. A sketch of the ideal apparatus is given in Figure \ref{fig:feed}.

We consider two diffusive channels realized by heterodyne or homodyne detectors of the emitted light with \emph{local oscillators} represented by the functions $h_j$:
\[
d=2, \qquad L_j(t)= \overline{h_j(t)}\, \alpha_j \sigma_-, \qquad \abs{h_j(t)}=1,
\quad \alpha_j\in \Cbb.
\]
The observation of the light in channel 1 will be used to control the stimulating laser light. The light in channel 2 is only observed, in order to analyse its properties. We are interested in controlling the properties of this part of the emitted light by controlling the atom via the stimulating laser; in particular we are interested in generating squeezed light.

In the case of \emph{homodyne detection} the local oscillator is fed by the light produced by the stimulating laser and we have $h_j(t)=\re^{-\ri \epsilon_j}\frac{f(t)}{\abs{f(t)}}$. This is the case illustrated in Figure \ref{fig:feed}.

In the case of \emph{heterodyne detection} the local oscillator is fed by an independent laser. For this light we could use for instance the so called phase diffusion model, \[
h_j(t)=\exp\left\{-\ri \epsilon_j-\ri \nu_j t-\ri k_{-j}B_{-j}(t)\right\}, \]
where $B_{-1}$ and $B_{-2}$ are extra noises (independent standard Wiener processes). The function $h_j$ represents a nearly monochromatic wave with a Lorentzian spectrum centred on $\nu_j$.

We introduce also four jump channels:
\[
d'=4, \quad L_k(t)= L_k, \quad \beta_k\in \Cbb, \quad \gamma>0, \quad \bar n\geq
0,\quad \sum_{i=1}^2 \abs{\alpha_i}^2+
\sum_{k=3}^4 \abs{\beta_k}^2=\gamma,
\]
\[
L_k=\beta_k \sigma_-, \quad k=3,4,5 \qquad
L_6=\beta_6 \sigma_+,  \qquad
\abs{\beta_5}^2=\abs{\beta_6}^2=\gamma \bar n.
\]
The jump channels 3 and 4 are electromagnetic channels:  channel 3 represents the emitted light reaching a photo-counter (\emph{direct detection}), while channel 4 represents the lost light, which is not observed. The output of channel 3 could be used again as a possible signal for closed loop control, but we use it here only to see properties of direct detection. As channel 2 is used to detect properties of the field quadratures (squeezing), the analogous channel 3 is used to explore the counting statistics of the emitted photons.

The counting channels 5 and 6 are used only to introduce dissipation due to a thermal bath; these channels are not connected to observation. So, channels 4, 5, 6 are introduced only to give dissipative effects in the system dynamics; to represent them in the jump form (as done here) or in the diffusive form has no influence on the physical quantities.

The stimulating laser light can be noisy and can be controlled by the output of the diffusive channel 1. In mathematical terms,
$f$ can be an adapted functional of $B_0$ (extra-noise) and $B_1$ (feedback of the output of the homodyne detector 1).

For this model we need a probability space $(\Omega,\Fscr,\Qbb)$ where $B_{-2},\ldots,B_2$ are Wiener processes, $N_3,\ldots,N_6$ are Poisson processes and they are all independent. According to the notations of Section \ref{sec:SME}, the filtration $(\Fscr_t)$ is generated by all these processes, while $(\Escr_t)$ is generated by $B_1$, $B_2$ and $N_3$.

From Eq.\ \eqref{randomLcal} and the assumptions of the model we get the random Liouville operator
\[
\Lcal(t)[\tau]=-\ri  \left[H_0+H_f(t),\, \tau \right] + \gamma \sigma_-\tau
\sigma_+-\frac \gamma 2 \left\{\sigma_+\sigma_-,\tau \right\}+ \gamma\bar n \left(\sigma_-\tau
\sigma_++\sigma_+\tau
\sigma_- - \tau\right) .
\]
Note that the whole of the randomness is in the wave $f(t)$ and it is due purely to noise in the laser light and to feedback, but this is enough to have a non Markovian evolution of the atom.

\subsection{The laser wave}

Let us denote by $f_0(t)$ the laser wave without feedback. We assume the laser to be not perfectly coherent and monochromatic and we describe this again by the phase diffusion model, i.e.
\begin{equation}\label{f0}
f_0(t)=\frac{\Omega_R} 2\, \re^{-\ri \left[\vartheta +\nu t + k_0 B_0(t)\right]}, \qquad
\Omega_R\geq 0,\quad \nu> 0, \quad k_0\in \Rbb,
\quad \vartheta \in (-\pi,\pi];
\end{equation}
$\Omega_R$ is known as \emph{Rabi frequency} and $\Delta\nu:=\nu_0-\nu$ as \emph{detuning}. It is easy to compute the spectrum of this wave; we get
\[
S_{f_0}(\mu):=\lim_{T\to +\infty} \frac 1 T \, \Ebb\bigg[\biggl| \int_0^T \re^{\ri \mu t}f_0(t)\rd t \biggr|^2\bigg]= \frac{\Omega_R^{\ 2}k_0^{\,2}}{k_0^{\,4}+4\left(\mu-\nu\right)^2}.
\]
So, $f_0$ represents a wave of constant amplitude $\abs{f_0(t)}^2=\Omega_R^{\ 2}/4$ and Lorentzian spectrum. The effect of such a stimulating laser wave in the heterodyne spectrum of a two-level atom has been studied in Refs.\ \cite{BarP02,Bar06}. In that papers, however, the presentation of the theory of continuous measurements was based on the use of Bose fields and quantum stochastic calculus.

The present theory gives also the possibility of describing feedback control, by assuming that $f(t)$ is a functional of the output current $J_1(s)$, $0\leq s \leq t$. We take $J_1$ to have the form given in Eqs.\ \eqref{Jl}, \eqref{a_ellF} with a very simple detector response function; to be concrete we take
\begin{equation}
J_1(t)= k \int_0^t\re^{-\frac \varkappa 2\left(t-s\right)} \rd B_1(s), \qquad k\in \Rbb, \quad \varkappa \geq 0.
\end{equation}
The functional dependence of $f$ on $J_1$ is determined by the implementation of the feedback mechanism, which should be chosen according to the aims of the controller and to the physical feasibility. Then, in principle, the quantities of interest could be computed by starting from simulations of the non linear SSE of  the model. However, in order to have an analytically computable spectrum and to have an idea of the possible behaviours of the model, we consider very simple forms for $f$.

A first choice is to modify the wave \eqref{f0} by a simple linear amplitude modulation and to take
\begin{equation}\label{fa}
f(t)=\frac{1} 2\, \re^{-\ri \left[\vartheta +\nu t + k_0 B_0(t)\right]}\left[\Omega_R +c\,\re^{\ri \vartheta_1}J_1(t-\delta)\right], \qquad
c\in \Rbb,
\quad \vartheta_1 \in (-\pi,\pi];
\end{equation}
$\delta\geq 0$ represents a possible delay. By taking $\delta=0$, $k=\sqrt{\varkappa/(2\pi)}$, $\varkappa \uparrow +\infty$ one formally obtains $J_1(t)=\dot B_1(t)$, the limiting case of the ``Markovian feedback'' of Wiseman and Milburn \cite{WisM93,WisM94}; in \cite{BarGL09} and in \cite[Chapt.\ 10]{BarG09} this feedback scheme has been formulated in terms of SMEs and applied to the homodyne spectrum of a two-level atom. The effect of delay in some models has been studied in \cite{WisM94,WisM10,CW11}.

A second choice of the control is to introduce a phase modulation of the wave \eqref{f0}. By taking this phase modulation linear in the output current we get
\begin{equation}\label{fp}
f(t)=\frac{\Omega_R} 2\, \re^{-\ri \left[\vartheta +\nu t + k_0 B_0(t)+ k_1 J_1(t-\delta)\right]},
\qquad \delta\geq 0, \quad k_1\in \Rbb.
\end{equation}
Our aim is only to show that it is possible to construct in a consistent way a theory with a non Markovian feedback and to demonstrate that, even in simple cases, the feedback can be used to control the squeezing. So, we simplify further on the expression \eqref{fp} and we take no delay, $\delta=0$, and $\varkappa=0$, $k=1$. Then, we have $J_1(t)=B_1(t)$; note that the formal instantaneous output is $\dot B_1(t)$ and, thus, $B_1(t)$ is the integrated current: $f(t)$ depends on the whole past of the signal detected by the homodyne detector 1. Our final choice for the laser wave is
\begin{equation}\label{fpp}
f(t)=\frac{\Omega_R} 2\, \re^{-\ri u(t)}, \qquad u(t)=\vartheta +\nu t + k_0 B_0(t)+ k_1 B_1(t).
\end{equation}

Whatever form we take for $f$, the feedback acts on the stimulating light and on the local oscillators, as illustrated by Figure \ref{fig:feed}. As $f$ enters the Hamiltonian \eqref{H(t)}, we can control to some extent the dynamics  of the atom. This possibility can be used to control the system state or the time to reach a final state \cite{WanW01,CW11}. Instead, what we want to study in this paper is how the feedback can be used to control the output of the quantum system, precisely the squeezing in channel 2 or the Mandel $Q$-parameter in channel 3.

\subsection{Control of the homodyne spectrum}

According to Eqs.\ \eqref{gen_sp}, to compute the spectrum of the light in channel 2 we need to compute first the quantities $n_2(t)$ and $d_2(t,s)$. The best way is to introduce a kind of stochastic rotating frame, that is to make a unitary transformation and to define the quantities
\[
\Lambda(t,s)[\tau]:=\re^{\frac\ri 2 \,u(t)\sigma_z}\Acal(t,s)\left[
\re^{-\frac\ri 2 \,u(s)\sigma_z}\tau \re^{\frac\ri 2 \,u(s)\sigma_z}\right]\re^{-\frac\ri 2 \,u(t)\sigma_z},
\]
where $\Acal(t,s)$ is the propagator defined at the end of Section \ref{sec:SME}.\ref{sec:lSME}, and
\begin{equation}\label{xi}
\xi(t):=\re^{\frac\ri 2 \,u(t)\sigma_z}\sigma(t)\re^{-\frac\ri 2 \,u(t)\sigma_z}=
\Lambda(t,0)\left[\tilde\rho_0\right],
\end{equation}
\[
\tilde\rho_0=\re^{\frac\ri 2 \,\vartheta\sigma_z}\rho_0\re^{-\frac\ri 2 \,\vartheta\sigma_z}.
\]
By using \eqref{lSME}, \eqref{fpp} and computing the stochastic differential of $\xi(t)$  we obtain
\begin{multline}\label{eq:xi}
\rd \xi(t) =\hat\Lcal(t)[  \xi(t_-)]\rd  t+ \sum_{i=0}^2 \Dcal_i(t)[
\xi(t_-)]\rd B_i(t)
\\ {}+
\sum_{k=3}^{6}\left(\frac 1 {\lambda_k}\,L_k   \xi(t_-) {L_k}^*
-  \xi(t_-)\right)\bigl( \rd  N_k(t) -\lambda_k\rd  t \bigr),
\end{multline}
\begin{multline}
\hat \Lcal(t)[\tau]=-\frac \ri 2 \left[\Delta\nu\sigma_z+\Omega_R \sigma_x,\, \tau \right] + \gamma \sigma_-\tau
\sigma_+-\frac \gamma 2 \left\{P_+,\tau \right\} + \frac {{k_0}^2+ {k_1}^2}4 \left(\sigma_z \tau \sigma_z
-\tau\right) \\ {}+ \ri k_1 \left( g_1(t)P_+\tau \sigma_+- \overline{g_1(t)}\, \sigma_- \tau P_+\right) + \gamma\bar n \left(\sigma_-\tau
\sigma_++\sigma_+\tau
\sigma_- - \tau\right) ,
\end{multline}
\[
\Dcal_0[\tau]= \frac \ri 2 \, k_0[\sigma_z,\tau],
\qquad \Dcal_1(t)[\tau]= \overline{g_1(t)}\, \sigma_- \tau + g_1(t)\tau \sigma_++ \frac \ri 2 \, k_1[\sigma_z,\tau],
\]
\[
\Dcal_2(t)[\tau]= \overline{g_2(t)}\, \sigma_- \tau + g_2(t)\tau \sigma_+, \qquad g_i(t)=\overline{\alpha_i}\,\re^{\ri u(t)}\, h_i(t), \qquad P_+=\sigma_+\sigma_-.
\]
By the link \eqref{xi} between $\xi$ and $\Lambda$, we have that also $\Lambda(t,s)$ satisfies Eq.\ \eqref{eq:xi} with initial condition $\Lambda(s,s)=\openone$.

Then, we get
\[
n_2(t)=\Ebb_{\Pbb^T_{\rho_0}}\left[m_2(t)\right]
=2\Real \Ebb_{\Qbb}\left[\Tr\left\{L_2(t)\sigma(t_-)\right\}\right]= \Ebb_{\Qbb}\left[\Tr\left\{\Dcal_2(t)\left[\xi(t_-)\right]\right\}\right],
\]
\[
d_2(t,s)=
1_{(0,+\infty)}(t-s)\Ebb_{\Qbb} \left[\Tr\left\{\Dcal_2(t)\left[ \Lambda(t_-,s)\left[\Dcal_2(s)\left[\xi(s_-)\right]
-n_\infty\xi(s_-)\right]\right]\right\}\right].
\]
All terms inside the expectation values are random and in general $n_2$ and $d_2$ can not be computed explicitly.

\subsubsection{Homodyning in channels 1 and 2}
In the case of homodyning as in Figure \ref{fig:feed}, we have $g_j(t)=\abs{\alpha_j}\re^{-\ri \vartheta_j}$, $j=1,2$, which gives $\Dcal_1$, $\Dcal_2$, $\hat\Lcal$ non random and independent of
time; the operators $L_k$ in \eqref{eq:xi} were already non random. This is the key point which makes the model computable. By Eq.\ \eqref{eq:xi}, $\Lambda(t_-,s)$ and $\xi(s_-)$ turn out to be $\Qbb$-independent and we get $\Ebb_\Qbb[\Lambda(t_-,s)]=\re^{\hat\Lcal (t-s)}$, $\Ebb_\Qbb[\xi(s_-)]=\re^{\hat\Lcal s}\left[ \tilde \rho_0\right]$. Finally, we get
\[
n_2(t)= 2 \abs{\alpha_2} \Real \Tr \left\{ \re^{-\ri \vartheta_2} \sigma_+ \re^{\hat \Lcal t}\left[ \tilde \rho_0 \right]\right\}, \qquad n_\infty=\lim_{t\to +\infty}n_2(t),
\]
\begin{multline*}
d_2(t,s)=
\Bigl[2\abs{\alpha_2}^2 \Real\Tr\left\{ \re^{-\ri \vartheta_2} \sigma_+ \re^{\hat \Lcal\left(t-s\right)}\left[\re^{\ri \vartheta_2} \sigma_- \re^{\hat \Lcal s}\left[ \tilde \rho_0 \right]
+ \mathrm{h.c.}\right]\right\}\\ {} -n_\infty n_2(t)\Bigr]1_{(0,+\infty)}(t-s) ,
\end{multline*}
where h.c. means ``hermitian conjugate term''. The modified Liouville operator $\hat \Lcal$ can be written in the Lindblad form as
\begin{multline}\label{Lnew}
\hat \Lcal[\tau]=- \ri  \left[\hat H,\, \tau \right] + \left(\gamma-\abs{\alpha_1}^2\right)\left( \sigma_-\tau
\sigma_+-\frac 1 2 \left\{P_+,\tau \right\} \right)+ K\tau K^*\\ {} - \frac 1 2 \left\{K^*K, \tau \right\}+ \frac {{k_0}^2}4 \left(\sigma_z \tau \sigma_z
-\tau\right)+ \gamma\bar n \left(\sigma_-\tau
\sigma_++\sigma_+\tau
\sigma_- - \tau\right) ,
\end{multline}
\[
\hat H=\frac 1 2 \left(\Delta\nu\sigma_z+\Omega_R \sigma_x\right)+\frac{k_1 \abs{\alpha_1}}4 \left( \re^{-\ri \vartheta_1}\sigma_+ +\re^{\ri \vartheta_1}\sigma_-\right),
\]
\[
K= \frac {\ri k_1}2\, \sigma_z + \abs{\alpha_1} \re^{\ri \vartheta_1}\sigma_-.
\]

By Bloch equation techniques we can compute the homodyne spectrum \eqref{gen_sp} of the light in
channel 2 \cite[Part II]{BarG09}. One can show that the Liouville operator \eqref{Lnew} has a unique equilibrium state and that, because of the limits contained in \eqref{gen_sp}, any dependence on the initial state $\rho_0$ disappears from the expression of the spectrum. The final result is $S_2(\mu)=S_{\mathrm{el}}(\mu)+ S_{\mathrm{inel}}(\mu)$,
\begin{equation}
S_{\mathrm{el}}(\mu)= 2\pi \abs{\alpha_2}^{2}v^2 \delta(\mu), \qquad v:= \cos \vartheta_2 \, d_1+
\sin \vartheta_2 \, d_2,
\end{equation}
\begin{equation}
S_{\mathrm{inel}}(\mu)= 1+\begin{pmatrix} \cos \vartheta_2 & \sin \vartheta_2& 0
\end{pmatrix}\frac {2\abs{\alpha_2}^2 } {A^2+ \mu^2}\left(A \begin{pmatrix} \cos \vartheta_2 (1+d_3)\\
\sin \vartheta_2 (1+d_3)\\ -v
\end{pmatrix}+v \vec u\right),
\end{equation}
\begin{equation}\label{matrixA}
A=\begin{pmatrix}\frac \Gamma 2 & \Delta\nu  & -k_1\abs{\alpha_1}\sin \vartheta_1 \\
-\Delta\nu  & \frac \Gamma 2 &  \Omega_R +k_1\abs{\alpha_1}\cos \vartheta_1\\
0 & -\Omega_R & \left(2\bar n +1\right)\gamma
\end{pmatrix}, \qquad \vec u = \begin{pmatrix} -k_1\abs{\alpha_1}\sin \vartheta_1 \\
k_1\abs{\alpha_1}\cos \vartheta_1 \\ \gamma\end{pmatrix},
\end{equation}
\begin{equation}\label{vec:d}
\vec d =-A^{-1}\vec u ,\qquad \Gamma=\left(2\bar n +1\right)\gamma+{k_0}^2 + {k_1}^2.
\end{equation}

Now we can study the dependence of this spectrum on our parameters, feedback
included, and learn how to control the emitted light. By certain choices of the
parameters one can obtain $S_{\mathrm{inel}}(\mu)<1$, which is interpreted as
squeezing of the light in the channel 2.

\begin{figure}[h]\begin{center}
\includegraphics*[scale=.40]{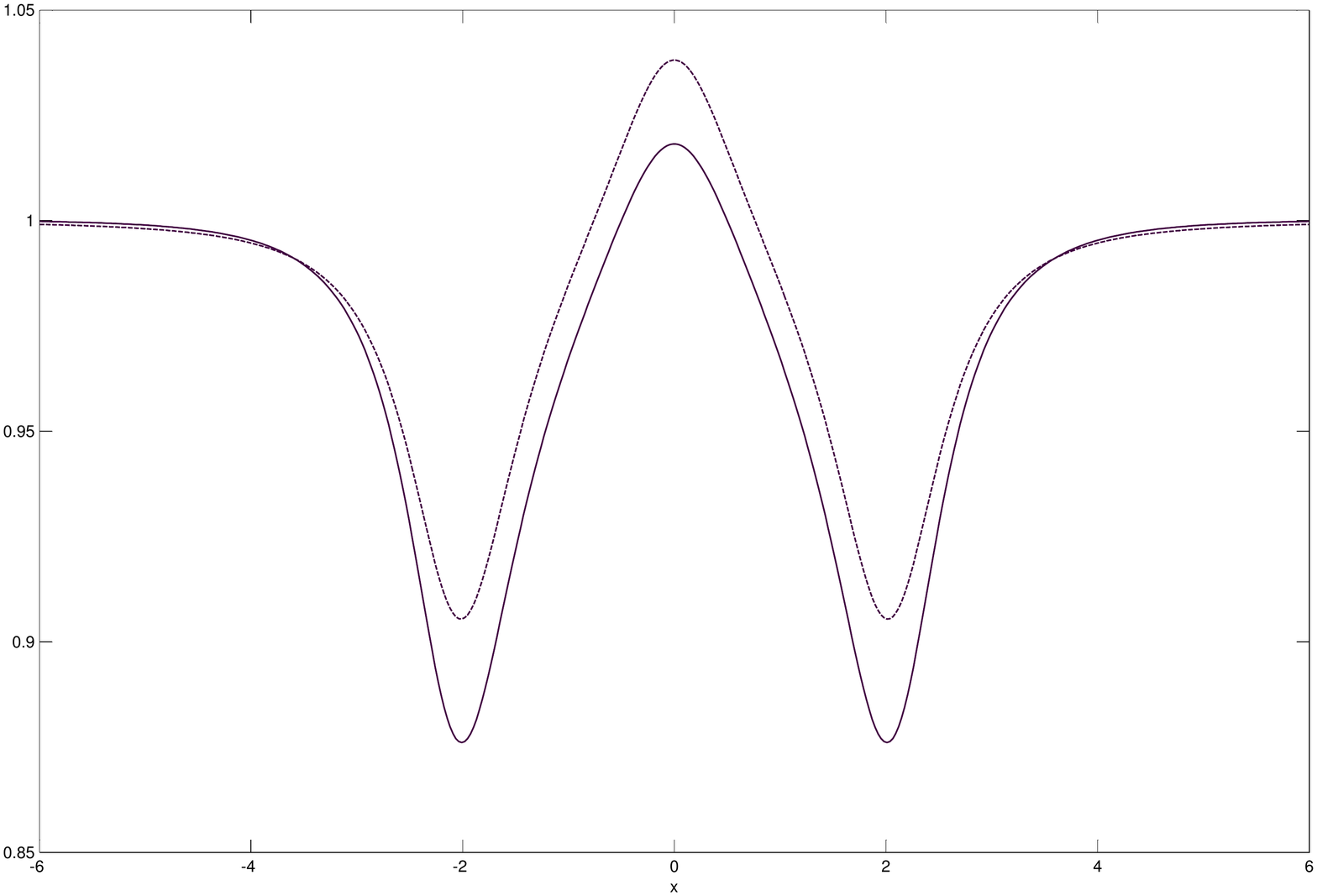}
\end{center}
\caption{Squeezing control. $S_{\mathrm{inel}}(\mu)$ with and without feedback for $\gamma=1$,
$k_0=\bar n=0$, $\abs{\alpha_1}^2=\abs{\alpha_2}^2=0.45$ and: (solid line)
$k_1=0.3213$, $\vartheta_1=-1.9307$, $\vartheta_2=-0.1540$, $\Delta\nu=1.3833$,
$\Omega_R=1.6150$; (dotted line) $k_1=0$, $\vartheta_2=-0.1784$,
$\Delta\nu=1.4937$, $\Omega_R=1.4360$.}\label{fig2}
\end{figure}

As an example, let us take  $k_0= 0$, $\gamma=1$, $\bar n= 0$, $\abs{\alpha_1}^2=\abs{\alpha_2}^2=0.45$. We choose the other parameters to have the deepest well for $S_{\mathrm{inel}}$ in a given point $\mu$. We denote by $s_{\mathrm{el}}=2\pi \abs{\alpha_2}^{2}v^2$ the coefficient of the delta-spike in the elastic part.
\begin{itemize}
\item For
$\Omega_R=0.366$, $\Delta\nu = 0$, $k_1=0.3371$, $\vartheta_1=-\pi$, $\vartheta_2=-1.5708$ we get $s_{\mathrm{el}}=1.0245$ and a minimum  in $\mu=0$ of $S_{\mathrm{inel}}(0)=0.8172$.
\item
For $\Omega_R=1.6150$, $\Delta\nu = 1.3833$, $k_1=0.3213$, $\vartheta_1=-1.9307$, $\vartheta_2=-0.1540$ we get $s_{\mathrm{el}}=1.4214$ and a minimum  in $\mu=2$ of $S_{\mathrm{inel}}(2)=0.8621$.
\item
For $\Omega_R=3.1708$, $\Delta\nu =2.5576$, $k_1=0.3249$, $\vartheta_1=-1.7863$, $\vartheta_2=-0.0760$ we get $s_{\mathrm{el}}=1.5356$ and a minimum in $\mu=4$ of $S_{\mathrm{inel}}(4)=0.8572$.
\end{itemize}
In Figure \ref{fig2} we plot the inelastic spectrum as a function of $x=\mu$ for a case with control and a case without control. The parameters are optimized to have the deepest minimum in $\mu=2$.

The values of the minima reported above are very similar to the ones found in \cite[p.\ 244]{BarG09}, where a very different scheme of feedback was considered (the instantaneous and singular feedback \`{a} la Wiseman-Milburn \cite{WisM93} discussed after Eq.\ \eqref{fa}). The two schemes are very different: one is a limiting case of amplitude modulation and the other an over-simplified case of phase modulation; it is not obvious a priori that one can obtain similar values for the optimal minima in the two cases. The difference of the two schemes is stressed also by the different values of the other control parameters which realize the minima in the two cases. For instance, the best minimum $S_{\mathrm{inel}}(4)=0.8572$ in $\mu=4$ in the present case is obtained for the Rabi frequency $\Omega_R=3.1708$ and the detuning $\Delta\nu =2.5576$, while in the case of \cite[Sect.\ 10.5.3]{BarG09} the best minimum $S_{\mathrm{inel}}(4)=0.8565$ is obtained for $\Omega_R=3.0721$ and $\Delta\nu =2.6310$. Without any feedback the best minimum is $S_{\mathrm{inel}}(4)=0.8830$ for $\Omega_R=2.9001$ and $\Delta\nu =2.8077$.

\begin{figure}[h]\begin{center}
\includegraphics*[scale=.40]{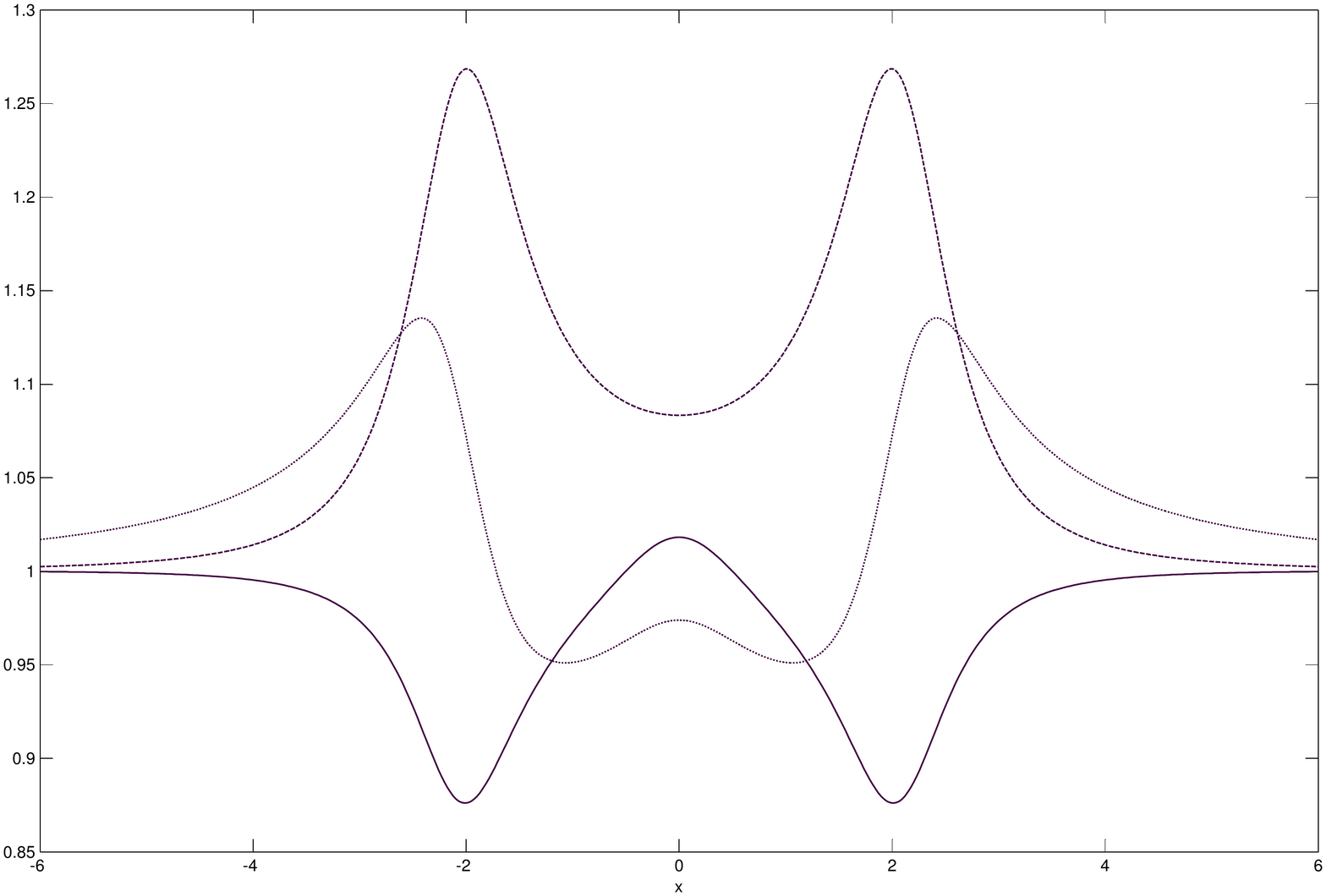}
\end{center}
\caption{Effect of Heisenberg uncertainty on $S_{\mathrm{inel}}(\mu)$. The parameters are $\gamma=1$,
$k_0=\bar n=0$, $\abs{\alpha_1}^2=\abs{\alpha_2}^2=0.45$,
$k_1=0.3213$, $\vartheta_1=-1.9307$, $\Delta\nu=1.3833$,
$\Omega_R=1.6150$ and: (solid line) $\vartheta_2=-0.1540$; (dotted line) $\vartheta_2=\frac \pi 4 -0.1540$; (dashed line) $\vartheta_2=\frac \pi 2 -0.1540$.}\label{fig3}
\end{figure}

By using quantum fields and quantum
stochastic calculus it is possible to prove that the Heisenberg uncertainty
principle implies that the product of $S_{\mathrm{inel}}(\mu)$, computed for a
certain value $\vartheta_2$, times the same quantity for $\vartheta_2+\pi/2$ is
always not less than 1 \cite{BarG08}:
\[
S_{\mathrm{inel}}(\mu;\vartheta_2)\,S_{\mathrm{inel}}(\mu;\vartheta_2+\pi/2)\geq1.
\]
In Figure \ref{fig3} we plot the previous case with feedback, taken with three different choices of $\vartheta_2$ in order to see the dependence on $\vartheta_2$, which changes the field quadrature under monitoring. We plot a case with a pronounced squeezing, the case with a $\pi/2$ shift in $\vartheta_2$ (anti-squeezing) and the case with a shift of $\pi/4$, just to see the behaviour of an intermediate field quadrature.

\subsection{Control of the photon statistics}
We maintain the same feedback as before, but now we study the Mandel $Q$-parameter of the direct detection $I_3=N_3$. We consider only the stationary regime at long times and we take
\[
Q_3(t)=\lim_{t_0\to+\infty}Q_3(t;t_0) =\lim_{t_0\to+\infty}\left(\frac {V_3(t;t_0)}{M_3(t;t_0)}-M_3(t;t_0)\right).
\]
By using again Bloch equation techniques to compute the expressions of
Section \ref{sec:SME}.\ref{sec:Qp}, we get
\[
\lim_{t_0\to+\infty}M_3(t;t_0)=\frac t 2 \abs{\beta_3}^2\left(1+d_3\right),
\]
\[
Q_3(t)=
\abs{\beta_3}^2\left[\left(\frac{1-\re^{-At}}{At}-1\right)A^{-1}\begin{pmatrix} d_1 \\ d_2 \\ 1+d_3 \end{pmatrix}\right]_3.
\]
where $\vec d$, $A$ are defined in Eqs.\ \eqref{matrixA}, \eqref{vec:d}. By taking  a long time interval we get
\[
Q_3:=\lim_{t\to+\infty}Q_3(t)=- \abs{\beta_3}^2
\left[A^{-1}\begin{pmatrix} d_1 \\ d_2 \\ 1+d_3 \end{pmatrix}\right]_3.
\]
According to the choice of the parameters we can obtain a sub-Poissonian or a
super-Poissonian $Q$-parameter: $k_0= 0$, $\gamma=1$, $\bar n= 0$,
$\abs{\alpha_1}^2=\abs{\beta_3}^2=0.45$,
\begin{itemize}
\item $\Delta\nu=0$, $k_1=1.0126$, $\Omega_R=1.0063$, $\vartheta_1=\pi$: $Q_3=-0.5094$;
\item $\Delta\nu=0$, $k_1=0$, $\Omega_R=0.7071$: $Q_3=-0.3375$;

\item $\Delta\nu=2$, $k_1=2.8515$, $\Omega_R=2.3516$, $\vartheta_1=2.6914$: $Q_3=-0.4356$;

\item $\Delta\nu=2$, $k_1=0$, $\Omega_R=2.9155$: $Q_3=0.0860$.
\end{itemize}
Moreover, one can see numerically that in the case of no feedback, $k_1=0$, and
$\Delta \nu=2$, we have $Q_3>0$ for all $\Omega_R>0$. So, in this case feedback
control is essential to get sub-Poissonian light.

Finally, let us observe that channels 2 and 3 are completely analogous; only the two detection apparatuses are different. What comes out from the numerical computations is that no relation seems to hold
between $Q_3<0$ and squeezing; with the parameters which give squeezing we have
$Q_3=0.0602$ (for $\Omega_R=1.6150$, $k_1=0.3213$, $\vartheta_1=-1.9307$,
$\Delta\nu = 1.3833$) and $Q_3=0.09508$ (for $\Omega_R=3.1708$, $k_1=0.3249$,
$\vartheta_1=-1.7863$, $\Delta\nu = 2.5576$).

\section{Conclusions}

In this article we have shown how the SSE or the SME with stochastic coefficients allows to model a wide class of non Markovian quantum evolutions in agreement with the axioms of quantum mechanics. Let us stress that the theory is fully ``quantum'' in that it gives completely positive evolutions and instruments and it is able to describe quantum effects as squeezing and anti-squeezing. This statement can be strengthened by showing, at least in the Markov case \cite{BarP96,Bar06}, the equivalence with a theory containing quantum systems, quantum fields, unitary evolutions and observation of compatible field observables.

What we have shown in the non Markov case is that our approach can describe several evolutions with memory effects, both in the case of open but unobserved systems and in the case of continuously measured systems.

The fundamental ingredient is to start from the \emph{linear} SSE \eqref{lSSE}: even if the coefficients are stochastic, starting from a linear equation for pure (unnormalized) states ensures a linear, trace preserving and completely positive mean evolution, which therefore can describe the evolution of an open system. Typically, such evolution is not Markovian, as the coefficients are stochastic, and therefore it does not satisfy a master equation, but it is defined by a SSE (or, equivalently, by a SME) plus an expectation.

Moreover, the unravelling of this evolution provided by the SSE and the SME can have a physical interpretation in the theory of continuous measurements. Besides the linearity of the SSE \eqref{lSSE} and the stochasticity of its coefficients, the other key ingredient is the introduction of the outputs processes $J_\ell$ and $I_h$ \eqref{outputs} and of their natural filtration $\Escr_t$ eventually smaller than the initial filtration $\Fscr_t$: combining a linear equation for pure (unnormalized) states, stochastic coefficients and conditional expectations gives instruments $\Ical_t$ which can model a non Markovian continuous measurement. Indeed, it is possible to describe not only several sources of classical noise, such as imperfections of the apparatus or coloured noises, but also the most general measurement based feedback. Thus we can handle also feedback schemes with some delay in the feedback loop. Again, both the evolutions of the a posteriori and a priori states do not satisfy directly closed equations, but they are defined by stochastic differential equations plus (conditional) expectations.

Moreover, as we are considering continuous measurements, our interest is not only for the evolution of the observed system, but also for the outputs of the observation: the outputs are quite general, they have both diffusive and counting components, and we study their distribution, providing useful formulae for the moments and the spectra of the diffusive outputs and for the Mandel $Q$-parameters of the counting outputs.

To present an application of the theory, we consider a two-level atom, stimulated by a non perfectly monochromatic and coherent laser, and we introduce a thermal bath, a continuous measurement by detecting its fluorescence light, and a closed loop control by modulating the stimulating laser according to the observed output. The resulting evolution of the atom is not Markovian because of the imperfections of the laser and because of the chosen feedback loop.

More precisely, the fluorescence light is divided in several channels where detectors are positioned; one of these performs homodyne detection and its output is used in the feedback loop with the aim of controlling, not the atom itself, but the light emitted in the other channels. The chosen feedback provides a non Markovian example where calculations can be explicitly performed up to find the homodyne spectrum and the Mandel $Q$-parameter of the fluorescence light and their dependence on the control parameters. Thus we study the role of the feedback loop in controlling the quantum properties of the emitted light, such as squeezing and sub-Poissonian statistics, we show that these quantum properties can be enhanced by feedback, but we find that there is no relation between them.

Even keeping our interest confined to the two-level atom, several further developments are possible, such as using the feedback here introduced to control properties of the atom itself or other properties of the emitted light, maybe local in time, or such as introducing other measurement and feedback schemes and then studying (numerically) the optimal one associated to a given target (e.g.\ ``maximal squeezing'' of the fluorescence light).


\begin{thebibliography}{99}
\bibitem{Car93} Carmichael, H. J. 1993 \textit{An Open System Approach to Quantum Optics}. Lectures
    Notes in Physics \textbf{m 18}. Berlin: Springer.
\bibitem{ZolG97}  Zoller, P.  \& Gardiner, C. W. 1997 Quantum noise in quantum optics: the
    stochastic Schr\"odinger equation. In \textit{Fluctuations quantiques, (Les Houches 1995)} (ed. S. Reynaud, E. Giacobino \& J. Zinn-Justin), pp. 79--136.  Amsterdam: North-Holland.
\bibitem{BreP02} Breuer, H.-P. \& Petruccione, F. 2002 \textit{The Theory of Open
    Quantum Systems}.  Oxford: Oxford University Press.
\bibitem{GarZ04} Gardiner, C. W. \& Zoller, P. 2004 \textit{Quantum Noise: A
    Handbook of Markovian and Non-Markovian Quantum Stochastic Methods with
    Applications to Quantum Optics}, Springer Series in Synergetics.
    Berlin: Springer.
\bibitem{Car08} Carmichael, H. J. 2008 \textit{Statistical Methods in Quantum
    Optics, Vol 2}. Berlin: Springer.
\bibitem{Bar90} Barchielli, A. 1990 Direct and heterodyne detection and other
    applications of quantum stochastic calculus to quantum optics. \emph{Quantum Opt.} \textbf{2},
    423--441.
\bibitem{BarB91} Barchielli, A. \& Belavkin, V. P. 1991 Measurements continuous in time and a
    posteriori states in quantum mechanics. \emph{J. Phys. A: Math. Gen.} \textbf{24},
    1495--1514.
\bibitem{Hol01} Holevo, A. S. 2001 \textit{Statistical Structure of Quantum Theory} Springer Lecture Notes     in Physics \textbf{m 67}. Berlin: Springer.
\bibitem{BarG09} Barchielli, A. \& Gregoratti, M. 2009 \textit{Quantum trajectories and
    measurements in continuous time: the diffusive case}. Lecture Notes in Physics \textbf{782}. Berlin: Springer.
\bibitem{DavL70} Davies, E. B. \& Lewis, J. T. 1970 An operational approach to quantum
    probability. \emph{Commun. Math. Phys.} \textbf{17}, 239--260.
\bibitem{Dav76} Davies, E. B. 1976 \textit{Quantum Theory of Open Systems}. London: Academic Press.
\bibitem{Oza84} Ozawa, M. 1984 Quantum measuring processes of continuous observables, J.
    Math. Phys. \textbf{25}, 79--87.
\bibitem{BarL85} Barchielli, A. \& Lupieri, G. 1985 Quantum stochastic calculus, operation valued stochastic processes and continual measurements in quantum
    mechanics. \emph{J. Math. Phys.} \textbf{26}, 2222--2230.
\bibitem{Bar86} Barchielli, A. 1986 Measurement theory and stochastic differential equations in quantum mechanics. \emph{Phys. Rev.} A \textbf{34}, 1642--1649.
\bibitem{Bel88} Belavkin, V. P. 1988 Nondemolition measurements, nonlinear filtering and
    dynamic programming of quantum stochastic processes. In \textit{Modelling and Control of Systems} (ed. A. Blaqui\`ere). Lecture Notes in Control and Information
    Sciences \textbf{121}, pp. 245--265. Berlin: Springer.
\bibitem{Bel89} Belavkin, V. P. 1989 A new wave equation for a continuous nondemolition
    measurement. \emph{Phys. Lett.} \textbf{A 140}, 355--358.
\bibitem{BarP96} Barchielli, A. \& Paganoni, A. M. 1996 Detection theory in quantum
    optics: stochastic representation. \emph{Quantum Semiclass. Opt.} \textbf{8}, 133--156.
\bibitem{Bar06} Barchielli, A. 2006 Continual measurements in quantum mechanics and
    quantum stochastic calculus. In \textit{Open quantum systems III} (ed. S. Attal, A. Joye \& C.-A. Pillet). Lecture Notes in Mathematics \textbf{1882}, pp. 207--291. Berlin: Springer.
\bibitem{DioS97} Di\'osi, L. \& Strunz, W.T. 1997 The non-Markovian
    stochastic Schr\"odinger equation for open systems. \emph{Phys. Lett. A}
    \textbf{235}, 569--573.
\bibitem{DioGS98} Di\'osi, L., Gisin, N. \& Strunz,  W.T. 1998 Non-Markovian
    quantum state diffusion. \emph{Phys. Rev. A} \textbf{58}, 1699--1712.

\bibitem{WisG08} Wiseman H. M.  \& Gambetta J. M. 2008 Pure-state quantum trajectories for general non-Markovian systems do not exist. \emph{Phys. Rev. Lett.} \textbf{101}, 140401.
\bibitem{BarH95} Barchielli, A. \& Holevo, A. S. 1995 Constructing quantum measurement
    processes via classical stochastic calculus. \emph{Stoch. Proc. Appl.} \textbf{58},
    293--317.
\bibitem{BarPP10} Barchielli, A., Pellegrini, C. \& Petruccione, F. 2010 Stochastic
    Schr\"odinger equations with coloured noise. \emph{EPL} \textbf{91}, 24001.
\bibitem{BarDPP11} Barchielli, A., Di Tella, P., Pellegrini, C. \& Petruccione, F. 2011
    Stochastic Schr\"odinger equations and memory.  In  \textit{Quantum probability and related topics}
    (ed. R.\ Rebolledo \&  M.\ Orszag), QP-PQ:
    Quantum Probability and White Noise Analysis \textbf{27},  pp.\ 52--67. Singapore: World Scientific.

\bibitem{BarP02} Barchielli, A. \& Pero, N. 2002 A quantum stochastic approach to the
    spectrum of a two-level atom. \emph{J. Opt. B: Quantum Semiclass. Opt.}
    \textbf{4}, 272--282.
\bibitem{Bel83} Belavkin, V. P. 1983 Theory of the control of observable quantum systems.
    \emph{Automat. Remote Control} \textbf{44}, 178--188.
\bibitem{WisM93} Wiseman, H. M. \& Milburn, G. J. 1993 Quantum theory of optical feedback via
    homodyne detection. \emph{Phys. Rev. Lett.} \textbf{70}, 548--551.
\bibitem{WisM94} Wiseman, H. M. \& Milburn, G. J. 1994 Squeezing via  feedback.
    \emph{Phys. Rev. A} \textbf{49}, 1350--1366.
\bibitem{GTV99} Giovannetti, V., Tombesi, P. \& Vitali, D. 1999 Non-Markovian quantum feedback from homodyne measurements: the effect of a non-zero feedback delay time. \emph{Phys. Rev.} A \textbf{60}, 1549--1561.
\bibitem{WanW01} Wang, J.  \& Wiseman, H. M. 2001 Feedback-stabilization of an arbitrary
    pure state of a two-level atom. \emph{Phys. Rev. A} \textbf{64}, 063810.
\bibitem{WanWM01} Wang, J., Wiseman, H. M. \& Milburn, G. J. 2001 Non-Markovian
    homodyne-mediated feedback on a two-level atom: a quantum trajectory treatment. \emph{Chem.
    Phys.}   \textbf{268},   221--235.
\bibitem{NKI09} Nishio, K., Kashima, K. \& Imura, J. 2009 Effects of time delay in feedback control of linear quantum systems. \emph{Phys. Rev.} A \textbf{79}, 062105.

\bibitem{WisM10} Wiseman H. M.  \& Milburn G. J. 2010 \textit{Quantum Measurement and Control}. Cambridge: Cambridge University Press.
\bibitem{BarG08} Barchielli, A. \& Gregoratti, M. 2008 Quantum continual measurements: the
    spectrum of the output. In \textit{Quantum probability and related topics} (ed. J. C. Garc\'{\i}a, R. Quezada \& S. B. Sontz). QP-PQ: Quantum Probability
    and White Noise Analysis, \textbf{23}, pp. 63--76. Singapore: World Scientific.
\bibitem{BarGL09} Barchielli, A., Gregoratti, M. \& Licciardo, M. 2009 Feedback control of
    the fluorescence light squeezing. 
    \emph{EPL} \textbf{85}, 14006.

\bibitem{BelE08} Belavkin, V. P. \& Edwards, E. 2008 Quantum filtering and optimal control. In \textit{Quantum Stochastic and Information} (ed. V.\ P.\ Belavkin \& M.\ Gu\c{t}\v{a}) pp.\
    143--205. Singapore: World Scientific.
\bibitem{BouVH08} Bouten, L. \& Van Handel, R. 2008 On the separation principle in quantum control. In \textit{Quantum Stochastic and Information} (ed. V.\ P.\ Belavkin \& M.\ Gu\c{t}\v{a}) pp.\
    206--238. Singapore: World Scientific.

\bibitem{Gough08} Gough, J. 2008 Optimal quantum feedback for canonical observables. In \textit{Quantum Stochastic and Information} (ed. V.\ P.\ Belavkin \& M.\ Gu\c{t}\v{a}) pp.\
    262--279. Singapore: World Scientific.
\bibitem{James08} James, M. R. 2008 Feedback control of quantum systems. In \textit{Quantum Stochastic and Information} (ed. V.\ P.\ Belavkin \& M.\ Gu\c{t}\v{a}) pp.\
    280--299. Singapore: World Scientific.

\bibitem{How02} Howard, R. M. 2002 \textit{Principles of random signal analysis and low
    noise design, the power spectral density and its applications}. New York: Wiley.

\bibitem{CW11} Combes, J. \& Wiseman, H. M. 2011
Quantum feedback for rapid state preparation in the presence of control imperfections.
\emph{J. Phys. B} \textbf{44}, 154008.

\end{thebibliography}
\end{document}